\documentclass[twocolumn,floatfix,prb]{revtex4}
\usepackage{graphicx} 	
\usepackage{bm} 		
\usepackage{amssymb}
\usepackage{amsmath}
\usepackage{amsfonts}
\usepackage{epstopdf}
\usepackage{mciteplus}

\begin{document}
\renewcommand{\section}[1]{{\par\it #1.---}\ignorespaces}
\newcommand{\swave}{$s$-wave}
\newcommand{\dwave}{$d$-wave}

\title{The possibility of measuring intrinsic electronic correlations in graphene using a d-wave contact Josephson junction}

\author{Annica M. Black-Schaffer}
 \affiliation{Department of Applied Physics, Stanford University, Stanford, California 94305}
 \author{Sebastian Doniach}
 \affiliation{Departments of Physics and Applied Physics, Stanford University, Stanford, California 94305}

\date{\today}
\begin{abstract}
While not widely recognized, electronic correlations might play an important role in graphene. Indeed, Pauling's resonance valence bond (RVB) theory\cite{Paulingbook} for the $p\pi$-bonded planar organic molecules, of which graphene is the infinite extension, already established the importance of the nearest neighbor spin-singlet bond (SB) state in these materials. However, despite the recent growth of interest in graphene\cite{Geim07, *CastroNeto09} since its isolation,\cite{Novoselov04} there is still no quantitative estimate of the effects of Coulomb repulsion in either undoped or doped graphene.
Here we use a tight-binding Bogoliubov-de Gennes (TB BdG) formalism to show that in unconventional \dwave\ contact graphene Josephson junctions the intrinsic SB correlations are strongly enhanced. We show on a striking effect of the SB correlations in both proximity effect and Josephson current as well as establishing a $1/(T-T_c)$ functional dependence for the superconducting decay length. Here $T_c$ is the superconducting transition temperature for the intrinsic SB correlations, which depends on both the effects of Coulomb repulsion and the doping level.
We therefore propose that \dwave\ contact graphene Josephson junctions will provide a promising experimental system for the measurement of the effective strength of intrinsic SB correlations in graphene.
\end{abstract}

\maketitle
Graphene is formed as a single layer of carbon atoms arranged in a honeycomb lattice with a peculiar band structure where the low energy excitations are massless Dirac quasiparticles which have a linear dispersion and a Fermi velocity $v_F \sim c/300$.
In comparison with quantum electrodynamics (QED), this low value for $v_F$ gives a large fine structure constant for graphene, $\alpha_g \sim 2$, and thus Coulomb interaction effects are expected to be strong. 
Recently this fact has been studied in terms of the possibility of Coulomb interactions driving undoped graphene insulating through correlations in the electron-hole channel,\cite{Khveshchenko01b, *Khveshchenko09, Drut09, *CastroNeto09view} but, in general, the effect of electronic interactions in graphene has mostly been ignored.
The standpoint was, however, quite different in the early RVB treatments of the $p\pi$-bonded planar organic molecules by Pauling\cite{Paulingbook} and others. Later, the RVB idea was also revived by Anderson\cite{Anderson87} as a mechanism for superconductivity in the high-$T_c$ cuprate superconductors.
Baskaran\cite{Baskaran02} proposed in 2002 a Hamiltonian which phenomenologically incorporates an effective $J {\bf S}_{\bf i} \cdot {\bf S}_{\bf j}$ term (${\bf S}_{\bf i}$ is the spin of atom ${\bf i}$, and ${\bf i}$ and ${\bf j}$ are nearest neighbors), ultimately derived from Coulomb interactions, as a model with which to estimate  the SB correlations in graphene. 
A few years ago, we studied the Cooper pairing channel of this Hamiltonian and showed that a superconducting gap with \dwave\ symmetry develops below $T_c$, where $T_c$ increases with the level of doping of the graphene.
 In the bulk, the \dwave\ state takes the form $d_{x^2-y^2} + id_{xy}$ and thus breaks time-reversal symmetry. 
Recently this state has been shown theoretically to survive on-site Coulomb repulsion\cite{Pathak08} as well as to emerge as an instability in a functional renormalization group flow,\cite{Honerkamp08} results that both further strengthen the argument that superconducting correlations are important in graphene.

There have been reports of granular superconductivity in graphite\cite{Kopelevich00, Kopelevich06inbook, Esquinazi08} as well as a possible interfacial superconducting state between graphite and sulfur in the graphite-sulfur composites.\cite{daSilva01, Moehlecke04, Kopelevich06inbook} Both of these might be experimental realizations of the intrinsic SB mechanism,\cite{Black-Schaffer07} but more experimental data is needed in order to establish such a conclusion.
From an experimental point of view, the main obstacle to reach the SB superconducting regime in graphene is that a very heavy doping is needed for an experimentally measurable $T_c$. The most frequently used method of applying a gate voltage does not nearly induce enough carriers into the graphene whereas chemical doping often faces significant material science problems and can also produce unwanted changes in other physical properties of the graphene. 
An exception to the latter possibly being sulfur doping\cite{Black-Schaffer07} which offers an explanation to the experimental results for the graphite-sulfur composites mentioned above. 
Another viable approach would be to search for systems where the SB pairing correlations are naturally enhanced. One obvious candidate would be superconducting graphene. Induced superconductivity was recently demonstrated in graphene superconductor-normal metal-superconductor (SNS) Josephson junctions manufactured by depositing superconducting contacts on top of the graphene.\cite{Heersche07} 
However, we recently showed that there is no enhancement of the SB correlations in conventional \swave\ contact junctions.\cite{Black-Schaffer09}
The primary source for this behavior is the incompatibility between the \dwave\ symmetric SB state and the \swave\ superconducting contacts. It is therefore logical to expand our previous study to include unconventional superconducting contacts where the symmetry of the contacts matches that of the SB superconducting state. 
Using the TB BdG formalism we will report below  a significant enhancement of the SB correlations in graphene \dwave\ contact SNS Josephson junctions.

\section{Method}
\begin{figure}[htb]
\includegraphics[scale = 0.6]{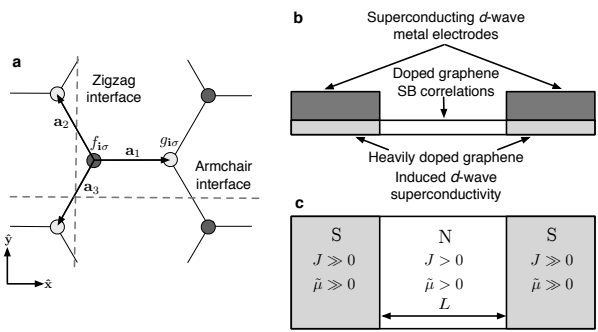}
\caption{\label{fig:junction} (a) The graphene honeycomb lattice. Schematic experimental setup (b) and model setup (c) for a \dwave\ contact graphene SNS Josephson junction.}
\end{figure}

We start with the phenomenological Hamiltonian for graphene proposed by Baskaran\cite{Baskaran02}:
\begin{align}
\label{eq:H_SB}
H =  & -t \!\! \! \! \sum_{<{\bf i},{\bf j}>,\sigma} \!\!\! (f_{{\bf i}\sigma}^\dagger g_{{\bf j}\sigma} + g_{{\bf i}\sigma}^\dagger f_{{\bf j}\sigma}) +
\sum_{{\bf i},\sigma} \mu({\bf i})(f_{{\bf i}\sigma}^\dagger f_{{\bf i}\sigma} + g_{{\bf i}\sigma}^\dagger g_{{\bf i}\sigma}) \nonumber \\
& -\sum_{<{\bf i},{\bf j}>}2J({\bf i})h_{\bf ij}^\dagger h_{\bf ij}.
\end{align}
Here $f_{{\bf i},\sigma}^\dagger$ and $g_{{\bf i} \sigma}^\dagger$ are the creation operators on the two inequivalent lattice sites of the honeycomb lattice, see Fig.~\ref{fig:junction}a, $t = 2.5$~eV is the hopping parameter, $<{\bf i},{\bf j}>$ indicates nearest neighbors, and $\mu$ is the effective chemical potential where $\mu = 0$ corresponds to the Dirac point. 
%
The SB correlations are modeled by the last term where $h_{\bf ij}^\dagger = (f_{{\bf i}\uparrow}^\dagger g_{{\bf j} \downarrow}^\dagger - f_{{\bf i}\downarrow}^\dagger g_{{\bf j} \uparrow}^\dagger)/\sqrt{2}$ is the spin-singlet creation operator on the bond between atoms ${\bf i}$ and ${\bf j}$.
Note that 
\begin{align}
\label{eq:SB}
-Jh_{\bf ij}^\dagger h_{\bf ij} \equiv J({\bf S}_{\bf i}\cdot {\bf S}_{\bf j} - \frac{1}{4}n_{\bf i}n_{\bf j})
\end{align}
and therefore the last term in equation (\ref{eq:H_SB}) both effectively models the SB correlations and shows its Cooper pairing nature.
The effective coupling constant $J$ has been estimated to be as large as $t$.\cite{Baskaran02}
Using $\Delta_{J{\bf a}}({\bf i}) = -\sqrt{2}J({\bf i})\langle h_{{\bf i},{\bf i+a}}\rangle$ as the mean-field order parameter, equation (\ref{eq:H_SB}) can solved with a double Bogoliubov-Valatin transformation. Here ${\bf a}$ denotes the nearest neighbor bond, see again Fig.~\ref{fig:junction}a, and we allow the order parameter to be independent on the three bonds.
The most favorable superconducting state in the bulk has $\Delta_{J{\bf a}} \propto (1,e^{i2\pi/3}, e^{i4\pi/3})$ which corresponds to a $d_{x^2-y^2}+id_{xy}$-wave symmetry in the Brillouin zone. 
As long as $J>0$ this state has a finite $T_c$ for any finite doping level and $T_c$ increases rapidly with either hole or electron doping.\cite{Black-Schaffer07}

In addition to equation~(\ref{eq:H_SB}), the influence of the superconducting contacts, deposited on top of the graphene sheet as shown in Fig.~\ref{fig:junction}b, has to be taken into account. We do this by assuming that the contacts induce into the S regions of the graphene a) an effective paring potential and b) a high doping level. 
For \dwave\ contacts the simplest effective pairing potential is the same $J$-term as for the intrinsic SB correlations. We therefore chose to model the effect of the contacts with a strong $J$-term in the S regions, see Fig.~\ref{fig:junction}c. The particular \dwave\ symmetry in the S regions is fixed to  the $d_{x^2-y^2}$-wave on the zigzag interface which is achieved by setting $\Delta_{J{\bf a}} \propto (2,-1,-1)$.\cite{Black-Schaffer07, Linder09} We have checked that a $90^\circ$ rotation of either interface or \dwave\ symmetry produces no notable change in the proximity effect. Note that the choice of \dwave\ symmetry is still unrestricted in the N region, though with the symmetry fixed in the S regions and relatively short junctions, the same symmetry choice in N is now energetically favored.
We should also note that we have set an artificially high value for $J$ in S in order to get a short superconducting coherence length $\xi$ and therefore reduce the computational size of the problem. However, we believe our conclusions are still applicable for an experimental setup with, e.g., a \dwave\ high-$T_c$ cuprate superconductor as the contact metal.
The doping is set to be high in the S regions since the superconducting contacts act as large electronic reservoirs directly in contact with the graphene. 
In the N region we assume that the doping level is finite in order to produce a non-zero $T_c$ for the intrinsic SB correlations. 
In fact, the doping in the N region is kept moderately high in our calculations in order to deal with numerically accessible $T_c$s for the SB superconductivity. But, since our results use only $T_c$ as the relevant SB parameter, this should not affect the generalization of our results to a more realistic lower doping regime.
We assume that the effective chemical potential abruptly drops at the SN interfaces. While this is a simplification, a moderate leakage of charge from S to N would only raise the intrinsic $T_c$ in the interface region and thus enhance the SB pairing effect.

The above framework can be solved self-consistently for the position-dependent $\Delta_{J{\bf a}}({\bf i})$ using the TB BdG formalism for ballistic graphene SNS Josephson junctions.\cite{Black-Schaffer08, Black-Schaffer09,Linder09}  We assume clean, smooth interfaces in order to reduce the computational effort but  we do not expect a moderately disordered interface to significantly change our results.\cite{Black-Schaffer09}
It is the use of an effective pairing potential instead of an effective order parameter in the S regions that allows us to solve self-consistently for the order parameter. This leads to a self-consistent calculation of the proximity effect depletion in S and leakage into N of the pair amplitude $F_{J{\bf a}} = \Delta_{J{\bf a}}/(\sqrt{2}J)$ and its norm $F$ as seen in Fig.~\ref{fig:JinNproxeff}. This should be compared to the analytical Dirac-BdG (DBdG) formalism developed for graphene SNS Josephson junctions\cite{Beenakker06, Titov06} where the order parameter is assumed to be constant in the S regions. Not only is a self-consistent solution necessary for studying the coupling between the externally induced \dwave\ state and the SB pairing correlations but a recent comparison between the two methods has also demonstrated the necessity for a self-consistent solution when unconventional contacts are considered.\cite{Linder09}

\section{Results}
\begin{figure}[htb]
\includegraphics[scale = 0.7]{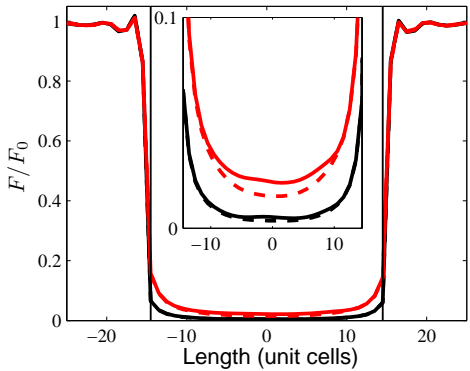}
\caption{\label{fig:JinNproxeff} Proximity effect in terms of the norm of the pair amplitude $F$ normalized to the bulk value in the contacts $F_0$ when SB correlations are ignored, i.e.~$J({\rm N}) = 0$ (black), and when SB correlations are present with $J({\rm N}) = t$ (red), where $t$ is the nearest neighbor hopping parameter.  The results are given at temperatures $T = 0.6 T_c$ (solid) and $T = 17 T_c$ (dashed) where $T_c$ is the superconducting transition temperature for the intrinsic SB correlations. Vertical lines indicate the SN interfaces. The junction length is $L = 30$~unit cells to be compared with the superconducting coherence length $\xi = 5$~unit cells in the contact regions. Inset shows a magnification over the N region.}
\end{figure}

The most straightforward physical quantity to study in the TB BdG framework is the proximity effect, or more specifically, the Cooper pair amplitude distribution in the SNS junction. If  a measurable signal of SB correlations were found in this quantity, there ought to be signs in other, more easily experimentally measurable, quantities such as the Josephson current. Fig.~\ref{fig:JinNproxeff} shows the norm of the pair amplitude, $F$, for a \dwave\ contact SNS junction.
%
In black are the results when the SB correlations are ignored at two different temperatures and, as seen, the results are essentially temperature independent in this range. When SB correlations are included (red) the pair amplitude in N increases. The pair amplitude inside N is in fact larger by almost an order of a magnitude than it would have been in a bulk sample with the same strength SB pairing. This means that in \dwave\ contact SNS junctions the intrinsic SB pairing is heavily benefitting from the proximity to the superconducting contacts. This is despite the large Fermi level mis-match (FLM) at the SN interfaces, due to different doping levels in S and N, which acts as an effective interface barrier.\cite{Linder09} 
However, since the extrinsic superconducting state and the intrinsic SB pairing here by design have the same symmetry this is probably not a too surprising result. More interesting is the temperature scale over which this enhanced SB pairing takes place. The solid curve is for $T = 0.6 T_c$, i.e.~when the doping level is sufficient for intrinsic SB superconductivity to be present, and the dashed curve is recorded when $T=17T_c$. We thus see a strongly enhanced proximity effect in the SNS junction even far above $T_c$. 

\begin{figure}[h!]
\includegraphics[scale = 0.7]{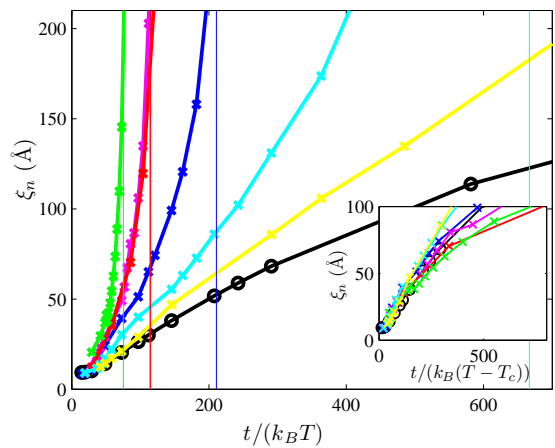}
\caption{\label{fig:decaylength} Superconducting decay length $\xi_n$ as a function of scaled inverse temperature $t/(k_B T)$ for $J({\rm N}) = 0$ (black, circles) and $J({\rm N})>0$ (colors, crosses) with increasing transition temperatures $T_c$ achieved by increasing $J({\rm N})$ (yellow, cyan, blue, red, green). Increasing the doping level in N instead (magenta) produces the same result. Vertical lines indicate the corresponding $T_c$s.
The inset shows the linear behavior of $\xi_n$ vs.~$t/(k_B(T-T_c))$ at moderately to large $(T-T_c)$.}
\end{figure}

One very closely related, but easier quantifiable, property to the proximity effect is the superconducting decay length, or normal state coherence length $\xi_n$, which is the length scale over which the pairing amplitude (exponentially) decay inside N:
\begin{align}
\label{eq:dlength}
F \propto e^{-x/\xi_n}.
\end{align}
Fig.~\ref{fig:decaylength} shows the inverse temperature dependence of $\xi_n$ for several different strengths of the intrinsic SB pairing. In black circles are the results when SB correlations are ignored whereas colored crosses show the results when they are present with the vertical lines indicating the corresponding intrinsic $T_c$. The red and magenta curves correspond to the same $T_c$ but achieved through different combinations of SB correlation strength $J$ and doping level, which demonstrates the dependence on $T_c$ only. 
%
From the figure we clearly see that to a good approximation $\xi_n \propto 1/T$ in the uncorrelated case. For any finite SB pairing inside N, $\xi_n$ will diverge at the corresponding intrinsic $T_c$ leading to a giant proximity effect for temperatures close to $T_c$. 
The same diverging behavior has recently also been predicted for the square lattice for both \swave\ and \dwave\ contact SNS junctions when the N region exhibits weak superconductivity of the same symmetry,\cite{Covaci06} although this study was conducted with no FLM at the SN interfaces.
The increase in proximity effect far above $T_c$ documented in Fig.~\ref{fig:JinNproxeff}, is here reflected in an increased decay length even at elevated temperatures. The inset in Fig.~\ref{fig:decaylength} shows that for temperatures moderately to high above $T_c$ we can extract  the following functional form for the superconducting decay length:
\begin{align}
\label{eq:Tdepxi}
\xi_n \propto \frac{1}{T-T_c}.
\end{align}
The limited accuracy due to finite model systems makes our numerical data deviate from this relationship at temperatures close to $T_c$, but we expect equation~(\ref{eq:Tdepxi}) to be valid even as $T \rightarrow T_c$.
From this we can conclude that any evidence of a divergence in $\xi_n$ at a non-zero temperature with a $1/(T-T_c)$ functional form in a \dwave\ contact graphene SNS junction would be a clear measurement of the strength of the intrinsic SB correlations in graphene.

\begin{figure}[tb]
\includegraphics[scale = 0.7]{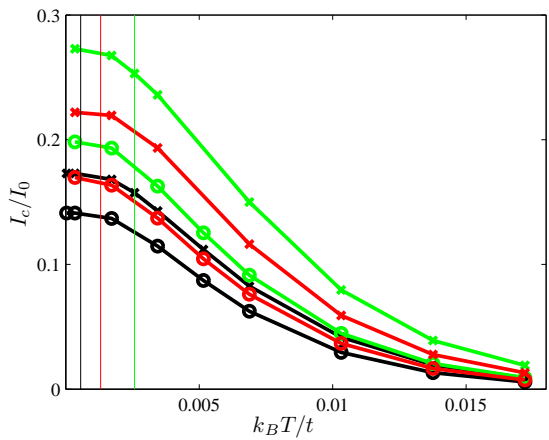}
\caption{\label{fig:IcvsT} Critical current $I_c$ in units of $I_0 = eW/(\hbar \xi)$ when $v_F$ is set to 1 for $J({\rm N}) = 0$ (circles) and $J({\rm N}) = t$ (crosses). Different colors represent increasing doping levels (black, red, green) in N. The corresponding $T_c$s are marked with vertical lines. The junction length is $L = 30$~unit cells to be compared to $\xi = 5$~unit cells.}
\end{figure}

It would also be useful to characterize the effect of SB correlations on the most accessible property of a SNS junction, namely the Josephson current. Fig.~\ref{fig:IcvsT} shows the critical current as a function of temperature when SB correlations are ignored (circles) and present (crosses) for three different doping levels in N. The corresponding $T_c$s for the SB pairing are again marked with vertical lines.
%
As seen, the Josephson current increases with decreasing temperatures in all cases. This is expected since $\xi_n$ is inversely proportional to the temperature. There is also a pronounced increase in the current when SB correlations are present in the junction, and, most importantly, the current is enhanced far above $T_c$. 
For systems with stronger SB correlations, the current increases more compared to the uncorrelated case at and below $T_c$. This effect is not due to the decay length per see, since $\xi_n$ diverges at $T_c$ for all strengths of the SB correlations, but related to the strength of the intrinsic SB superconducting state in N.

\section{Discussion}
The results in Figs.~\ref{fig:JinNproxeff}-\ref{fig:IcvsT} clearly demonstrate that the intrinsic SB correlations in graphene are greatly enhanced in a graphene \dwave\ contact SNS Josephson junction even far above their intrinsic $T_c$.
Since the intrinsic SB correlations at the currently experimentally fairly limited doping levels of graphene are expected to be rather weak, this enhancement even far above $T_c$ is crucial for any experimental detection and a finding of a finite $T_c$ in a graphene \dwave\ contact SNS Josephson junction will directly establish the strength of the effective SB coupling $J$ as a function of the doping level of the graphene.

%
A.M.B.-S. acknowledges partial support from the Department of Applied Physics and the School of Humanities and Sciences at Stanford University.

\bibliographystyle{apsrevM}
\ifx\mcitethebibliography\mciteundefinedmacro
\PackageError{apsrevM.bst}{mciteplus.sty has not been loaded}
{This bibstyle requires the use of the mciteplus package.}\fi

\end{document}